\documentclass[runningheads,a4paper]{llncs}
\usepackage{eso-pic}
\AddToShipoutPicture*{
\put(133,160){
\scriptsize
Proceedings of WLPE 2013, {\tt arXiv:1308.2055}, August 2013.
}}
\usepackage{times}
\usepackage{url}
\usepackage{fancyvrb}
\usepackage{color}
\usepackage{ifthen}
\usepackage{calc}
\begin{document}

\title{Why It's Nice to be Quoted: Quasiquoting for Prolog}

\author{Jan Wielemaker\inst{1} \and Michael Hendricks\inst{2}}
\institute{Web and Media group,
	   VU University Amsterdam, \\
	   De Boelelaan 1081a, 1081 HV Amsterdam, The Netherlands, \\
	   \email{J.Wielemaker@vu.nl}
	   \and
	   Hendricks Solutions, LLC, Hanna, Wyoming, USA, \\
	   \email{michael@ndrix.org}}

\maketitle
\bgroup
\input{swipl.sty}

\begin{abstract}
Prolog's support for dynamic programming, meta programming and text
processing using context free grammars make the language highly suitable
for defining \textit{domain specific languages} (DSL) as well as
analysing, refactoring or generating expression states in other
(programming) languages. Well known DSLs are the DCG (Definite Clause
Grammar) notation and constraint languages such as CHR. These extensions
use Prolog operator declarations and the \mbox{\{\ldots\}} notation to
realise a good syntax. When external languages, such as HTML, SQL or
JavaScript enter the picture, operators no longer satisfy for embedding
snippets of these languages into a Prolog source file.  In addition,
Prolog has poor support for quoting long text fragments.

Haskell introduced \textit{quasi quotations} to resolve this problem.
In this paper we `ported' the Haskell mechanism for quasi quoting to
Prolog.  We show that this can be done cleanly and that quasi quoting
can solve the above mentioned problems.
\end{abstract}

%\tableofcontents

%================================================================
\section{Introduction}
\label{sec:intro}

Prolog is commonly used for tasks where it needs to manage snippets of
code written in the syntax of an external language such as HTML, SQL or
JavaScript. Such code snippets often do not comply with the standard
Prolog syntax for various reasons \cite{WLPE/Wielemaker/2012}:

\begin{itemize}
    \item
Using operator syntax, with only prefix, infix and postfix
operators are insufficient.  For example, consider the JavaScript
\texttt{case} statement.
    \item
Basic lexical primitives are incompatible.   For example, consider textual
content in HTML documents.
    \item
The Prolog term that results from parsing an expression is
insufficient for recovering the intent in the target language.
For example, consider identifier names that start with an uppercase
letter, which will read as a Prolog variable.
\end{itemize}

In \cite{WLPE/Wielemaker/2012}, we claim that some simple languages can
be covered well by tweaking the Prolog syntax using operators, while it
is possible to realise acceptable representations for some other, e.g.,
XML based languages by using a Prolog syntax to represent the (simple)
data model of these language. In other cases, embedding of external
languages can be supported partially using semantic transformations. For
example, calls to Prolog predicates whose functor matches a table in an
RDBMS and whose arity matches the number of columns can be translated
into SQL queries \cite{Draxler91}. In \cite{WLPE/Wielemaker/2012}, we
propose a number of syntactic extensions to broaden the range of
languages that can be supported using one of the above means.

In this paper, we propose a solution for complex external languages for
which none of the above achieves a satisfactory embedding. The solution
is called \jargon{quasi quotations}, where we borrow the integration
into the language from Haskell \cite{DBLP:conf/haskell/Mainland07}. A
Haskell quasi quotation, as a syntactic element, has a syntax identifier
(a function symbol) and a snippet of quoted text. The function
associated with the syntax identifier is called during parsing and can
manipulate the text and massage it to fit further processing by Haskell.
It is good practice for the function to create an abstract syntax tree
by parsing the text according to the rules of the external language. As
we will see in this paper, more lightweight approaches can also be
sufficient.

This paper is organised as follows. After examining related work, we
revisit the problem statement and describe why current support in Prolog
is insufficient to solve this problem in a satisfactory way. In
\secref{quasiquote} we introduce the notion of quasi quoting and how it
fits into managing strings that present data structures from other
languages. In \secref{plqq}, we introduce the concrete solution proposed
for SWI-Prolog. This is followed by examples, future work and the
conclusions.

\section{Related work}

Quasi quotations are used for embedding snippets in many `scripting'
languages because they form an easy to understand mechanism to create
long strings or ASTs (\jargon{Abstract Syntax Tree}) with embedded
variables that are given values determined by the environment. Pure
textual replacement carries the risk of \jargon{injection} attacks,
which triggered research towards more safe \jargon{template} engines.%
\footnote{
TAL (\url{http://www.owlfish.com/software/simpleTAL/tal-guide.html})\\
Cheetah (\url{http://www.cheetahtemplate.org/examples.html})\\
Django (\url{https://docs.djangoproject.com/en/dev/ref/templates/})\\
Jinja2 (\url{http://jinja.pocoo.org/docs/templates/}).}
Notably modern scripting languages such as Python and Ruby provide a
rich pallete of such techniques. This paper follows the reasoning and
technique described in ``Why It's Nice to be Quoted: Quasiquoting for
Haskell'' \cite{DBLP:conf/haskell/Mainland07}. We refer to this article
for further background information. Another interesting system is Camlp4
\cite{Donham:2010:CTH:1900160.1900167}, which provides parsing and `anti
parsing' infrastructure for OCaml. Both approaches allows for extending
the host language. The main difference with our approach is that we
merely propose a hook into the Prolog parser that enable smart
processing of quasi quotations without dictating how these quotations
are processed.

\section{The need to embed snippets of external languages in Prolog}
\label{sec:needqq}

We will explain this need by an example from SWI-Prolog's web page
generating facilities \cite{TPLP06}. Web pages are by no means the only
place where these problems arise \cite{DBLP:conf/haskell/Mainland07}, but the
domain is quite familiar to many programmers and is of particular
interest to us because generating web pages is an important application
area for SWI-Prolog. A modern web page typically consists of HTML that
creates the initial DOM structure, CSS that provides styling and
JavaScript for creating interactive components. Our solution for
generating HTML is similar to PiLLoW \cite{PiLLoW} and based on
representing the HTML DOM structure as a nested Prolog term.  The
SWI-Prolog HTML page generation library provides good solutions for

\begin{itemize}
    \item Generating the page structure.
    \item Definition and reuse of DCGs that generate page fragments.
    \item Create references (URLs) to other pages on the server.
    \item Modular inclusion of resources, such as style and
          JavaScript files including dependency tracking and
	  automatic ordering by computing a partial order based
	  on the dependencies.
\end{itemize}

Pages generated with this infrastructure are guaranteed to be
syntactically correct, regardless of the data from which the page is
created. This built-in protection against injection attacks is
considered a key asset of the server infrastructure and a property that
we wish to maintain when integrating JavaScript into the page.

In earlier publications \cite{WLPE/Wielemaker/2012,TPLP06} we already
identified two missing pieces: (1) longer (HTML) text fragments and (2)
JavaScript fragments. The first problem can be solved using
PWP\footnote{Prolog Well-formed Pages,
\url{http://www.swi-prolog.org/pldoc/package/sgml.html}}. PWP was
developed by Richard O'Keefe and can be considered the opposite of the
above described Prolog based page generator. The page is described in
XML and reserved XML elements and tags trigger the generation of dynamic
page components based on the evaluation of Prolog goals. Like the Prolog
page generator, PWP protects against injection attacks because it
considers solutions from Prolog as data that is used to extend the XML
DOM, which is subsequently serialized for generating the final page. PWP
simplifies the specification of pages with large amounts of text, but
the rule format is less natural and it lacks the modularity of the
Prolog based generator.

We have been struggling with JavaScript, trying various approaches.
Currently, there are a large number of JavaScript frameworks that
provide `widgets' for use on web pages. Popular examples are
YUI\footnote{\url{http://yuilibrary.com/}} and
jQuery\footnote{\url{http://jquery.com/}}. Including these widgets
typically demands creating a DOM structure that carries an
\verb"id" attribute and represents the data (e.g., a menu may be
represented as a \verb"ul" list or an advanced text editor can be
represented as a \verb"textarea"). This DOM element is transformed into
the target widget by calling a JavaScript initialisation function that
modifies the DOM of the target element and installs \textit{event
handlers} that provide the dynamic behavior of the element. The
initialisation function often takes a \textit{configuration} object (a
JavaScript \jargon{object literal}, \mbox{\{name:value, \ldots\}}),
where some of the values are \textit{callback functions}, represented as
JavaScript lambda functions.

The HTML framework is well equipped to create the initial DOM and pull
in the JavaScript and CSS resources using its dependency framework. The
initialisation call is easy enough to abstract, except when the
configuration object contains lambda functions. These are often required
to massage JSON data from the server into the desired visualisation and
support AJAX based operations, such as computing auto-completion
candidates. It is quite common for such objects to require dozens of
lines of JavaScript. Representing JavaScript lambda functions requires
support for the complete complexity of the language. One way to do
this is to represent the JavaScript fragment as a list of literals,
mixed with DCG rules that insert content generated from Prolog, such as
URLs or JavaScript literal objects. This leads to code as below, which
we consider hard to type on a keyboard, hard to read and, because it is
so hard to read, often subject to syntax errors. Finally, the code below
is subject to injection attacks, unless we hand the variables \arg{URL}
and \arg{Id} to a grammar that generates valid JavaScript string content
from any Prolog atom.

\begin{code}
        [ '$.ajax({ url: "',URL,'",\n',
          '         data: { id: "',Id,'",\n',
          '               }\n',
          '       });\n'
        ]
\end{code}

\noindent
We tried several designs to improve on the above, none of which we
considered satisfactory. Below are the main directions that we tried.

\begin{itemize}
    \item Abstract away.  This implies using the HTML page generation
          facilities to generate e.g., make a JavaScript call,
	  initialize a variable from Prolog data, etc.  To remain
	  managable, a fairly high level of abstraction is needed
	  that is geared towards the JavaScript framework used.  This is
	  problematic because it makes it hard for the programmer to
	  relate the JavaScript examples from the framework
	  documentation to the Prolog code.  Although it is possible
	  to abstract some of the lambda functions, there is
	  too much variation to deal with all of them.
    \item Put the JavaScript in a separate file.  This creates many
          short files that typically only support a specific generated
	  HTML page because the details of the required JavaScript, such
	  as locations on the server that must be addressed vary from
	  page to page.  Because there is no formal relation between
	  the two pages, it is hard to relate them and keep them in
	  sync.
    \item Create application specific JavaScript resources that can be
	  configured without the need for lambda functions in the
	  configuration object. The problem here is that the
	  application developers create their own refinement of the
	  external widget library that requires understanding and
	  documentation, while the externally provided libraries are
	  already quite high level. In other words, the newly created
	  layer adds mostly new complications for managing and
	  understanding the code.
\end{itemize}

We have come to the conclusion that (1) we need a mechanism that allows
for including JavaScript into the Prolog page generation source code,
(2) the representation of JavaScript in the Prolog source must be easy
to type on a keyboard and understand, (3) the generated JavaScript
should be safeguarded against injection attacks without explicit calls
to encode data and (4) syntactic validation of the generated fragments
are likely to improve productivity.

\section{Using Prolog syntax}
\label{sec:syntaxext}

First, we investigated to what extent the syntax extensions proposed in
\cite{WLPE/Wielemaker/2012} would help to represent JavaScript naturally
using Prolog syntax. We realised support for the empty argument lists
(e.g., \verb"name()"), array notation and function bodies using
\mbox{\{\ldots\}}. These were resolved after a proposal by Jose Morales,
which extends the notion of operators to lists and curly-bracket blocks.
In addition to our hope of improving JavaScript support, list
subscription and curly-bracket attribute lists are in use with B-Prolog
and ECLiPSe.

We have realised a prototype that can express a fair deal of the
JavaScript syntax. However, the following problems remain unresolved:

\begin{itemize}
    \item It requires ! and . to be defined as operators.  These
          operators are known to cause ambiguity issues.
    \item Quoting of identifiers may be needed (e.g., \verb"'String'(...)")
    \item A special symbol is needed to distinguish identifiers from strings.
          ("string" cannot be used because it is a list of integers).
    \item Restrictions are needed with regard to spaces.  No space is
          allowed after function symbols and keywords such as
	  \const{while}.  In some places additional spaces are required
	  to prevent Prolog from reading two JavaScript tokens as a
	  single Prolog token.
    \item Some symbols clash with Prolog.  Consider \chr{\Sbar} or \chr{\Spercent},
          which play a totally different role in the Prolog syntax.
    \item Some JavaScript snippets can be turned into valid Prolog
          syntax, but the resulting AST is ambiguous. For example,
          \verb"++a" is different from \verb"a++", but both result in
	  the Prolog term \verb"++(a)".
\end{itemize}

We believe that the result would have been more usable than using Prolog
quoted atoms for short (1-10 lines) JavaScript snippets, that include
simple lambda functions such as making trivial AJAX callbacks. Our main
point of doubt is that it is hard to convey the restrictions to casual
Prolog users, which makes it likely to get surprised by output that
differs from the expectations and can only be understood through deep
knowledge of Prolog's syntax and underlying term representation.

\section{The three problems}
\label{sec:thebigthree}

If we cannot include JavaScript using Prolog syntax with user-defined
operators, the other option is to use (quoted) text. For this, we need
to solve three orthogonal problems:

\begin{enumerate}
    \item Allow for long quoted text.  Current (ISO) Prolog only provides
    single and double quoted text, which follow the same rules. In
    particular

    \begin{itemize}
        \item Quoted text cannot span more than one line, unless the
	      newline is escaped with a backslash.
        \item The backslash and quote must be escaped with a backslash.
	\item If good layout of the output is desirable, there is
	      no way to indent continuation lines in the Prolog source
	      such that the indentation does not show up in the output.
    \end{itemize}

    With these restrictions, simple copy and paste of example code
    must be followed by a tedious process of making the text fit the
    Prolog syntax, while the reverse is needed to test the code, for
    example in a JavaScript console.

    \item Relate Prolog variables (data) to references in the quoted
    material.

    \item Establish a \emph{safe} way to embed data into the template string.
\end{enumerate}

These problems are independent.  We illustrate this with a small HTML
example that presents a page with the current time.  We use an HTML
skeleton to make the comparison fair.  In pure Prolog, this could be
achieved using the following code snippet:

\begin{code}
        ...,
        get_time(Now),
        format_time(atom(Date), '%+', Now),
        print_html([date=Date],
                   '<h1>My digital clock</h1>\n\
                    <p>It is now {{date}}</p>').
\end{code}

\noindent
The task of \index{print_html/2}\predref{print_html}{2} is to relate the template variable \const{date}
to the Prolog variable \arg{Date} (third item above), determine the
lexical context of \mbox{\{\{date\}\}} to know the proper encoding that
needs to be done, realise this encoding on the Prolog atom and create
the proper output string. This is the second item on the list above.
Ideally, we would like an error if the HTML fragment is malformed to
begin with.

In pure ISO Prolog, all work must be done at runtime. Many Prolog
systems provide \index{goal_expansion/2}\predref{goal_expansion}{2} or similar, which allows for doing the
analysis of the template at compile time. What remains on our wishlist
is to (1) get rid of \verb"[date=Date]" and (2) get rid of complicated
escape sequences for languages that frequently use quotes or
backslashes. We propose to solve both issues using quasi quotations for
Prolog.

\section{Quasi Quotations}
\label{sec:quasiquote}

Quasi quotations find their origin in
linguistics\footnote{\url{http://en.wikipedia.org/wiki/Quasi-quotation}}
and introduces \textit{variables} into textual expressions. They are
commonly used in \textit{scripting} languages. For example:

\begin{code}
a=world; echo "Hello $a"        // Unix shell
$a = "world"; print "Hello $a"; // Perl
\end{code}

\noindent
This approach is natural to the (novice) programmer. Unfortunately it
does not work for Prolog because nothing in the normal Prolog
compilation process provides access to the names of variables. In
addition, plain text insertion is a direct invitation to
\textit{injection} attacks. As described in \secref{thebigthree}, the
template replacement code must be aware of the syntax to perform a safe,
i.e., properly escaped, insertion of the variable.

Haskell quasi quotes resolve the injection problem (which is described
as a \emph{typing} problem in \cite{DBLP:conf/haskell/Mainland07}) by
associating a syntax identifier directly with the quoted data. The
syntax identifier is associated with a function which typically parses
the quoted material into an abstract syntax tree that can represent the
target language.

\section{Quasi Quotations in Prolog}
\label{sec:plqq}

Realising quasi quotations requires for a syntactic construct that (1)
provides long quoted strings, (2) associates the quotation to a
predicate that can act on it according to the requirements of processing
the external language and (3) provides access to the clause's variable
dictionary.

There are few options for adding a syntax extension to Prolog because
Prolog `symbol' characters glue together to form an atom to which
operator properties can be assigned. According to Ulrich
Neumerkel,\footnote{\url{https://lists.iai.uni-bonn.de/pipermail/swi-prolog/2013/010422.html}}
taking sequences of Prolog solo character that do not form names is a
good starting point. Among others, this allows for \verb"{|...|}".
Combined with a term that identifies the syntax, we propose the
following syntax for Prolog quasi quotations, where \arg{Syntax} is an
arbitrary (callable) Prolog term:

\begin{quote}\
	\qqopen
	\textit{syntax-identifier}
	\qqsep
	\textit{quoted-material}
	\qqclose
\end{quote}

In the rather unlikely event that \qqclose{} needs to be embedded in
the quoted material, this can be realised in two ways: (1) define the
predicate that processes the quotation to respect some escape sequence
or (2) use the existing flexibility of the target language to avoid
\qqclose{}. An example of (1) could be to introduce the mapping
\verb"\\" $\rightarrow$ \verb"\",
\verb"\{" $\rightarrow$ \verb"{",
\verb"\}" $\rightarrow$ \verb"}", after which the user can write
\verb"|\}".  An example of (2) can be to insert a space between the
two if this does not change the semantics or write \verb"\u007c}"
inside a string if the target language supports \verb"\uXXXX" escapes
inside strings.

Orthogonal to the syntax is the mapping of the quoted material to a
Prolog term, which can be a full AST of the snippet or a simplified
representation as used in \secref{jsqq}. This mapping is defined by the
\textit{syntax-identifier}, a callable term to which we refer as
\arg{Syntax} from now on. At the same time, quasi quotation merges
(Prolog) variables from the environment into the quoted material. Quasi
quotations are processed as follows:

\begin{itemize}
    \item The predicate \index{read/1}\predref{read}{1} is modified to recognize the \arg{Syntax}
          term and the quoted text.
    \item After all normal processing is finished, \index{read/1}\predref{read}{1} performs
	  the following call:

\bgroup\footnotesize
\begin{code}
call(+SyntaxName, +Content, +SyntaxArgs, +VarDict, -Result)
\end{code}

\noindent
\egroup

	  Here, \arg{SyntaxName} is the functor of the \arg{Syntax} term
	  and \arg{SyntaxArgs} is the list of arguments, i.e.,
	  \verb"Syntax =.. [SyntaxName|SyntaxArgs]".  In \secref{examples},
	  we will see why it is useful to split the functor from the
	  arguments.  \arg{Content} is an opaque handle to the quoted
	  material as we will see later.  \arg{VarDict} is
	  \mbox{\arg{Name} = \arg{Var}} list conforming to \index{read_term/3}\predref{read_term}{3}
	  and \index{write_term/3}\predref{write_term}{3}.  \arg{Result} is determined by the call to
	  the deterministic predicate \arg{SyntaxName}/4.  The predicate
	  \index{read/1}\predref{read}{1} inserts \arg{Result} at the location of the quasi
	  quotation in the output term.
    \item The \arg{Result} term (and thus the quasi quotation) must be
	  a goal (see \secref{sparqlqq}) or appear as the argument of
	  a goal that processes (often serializes) the result (often
	  an AST).  For example, \index{reply_html_page/3}\predref{reply_html_page}{3} in \secref{htmlqq}
	  is designed to serialize the HTML DOM (AST) produced by the
	  HTML parser called by the quasi quoter.
\end{itemize}

We provide two support predicates for \arg{SyntaxName}/4 to process the
result.  Note that there are no restrictions on how the called predicate
combines the quoted text with the syntax arguments and variable
dictionary to construct the final term.

\begin{description}
    \predicate{phrase_from_quasi_quotation}{2}{:Grammar, +Content}
Calls the grammar \arg{Grammar} on the list formed by \arg{Content}.
This predicate uses the \jargon{pure input} library described in
\cite{CICLOPS/Wielemaker/2008} to parse the content.  Syntax errors
may be raised using the non-terminal \index{syntax_error//1}\dcgref{syntax_error}{1}, which produces
a precise syntax location that consists of the file, line number,
line position and character count.

    \predicate{with_quasi_quotation_input}{3}{+Content, -Stream, :Goal}
Calls \arg{Goal} on the Prolog stream \arg{Stream}.  The stream position
information reflects the location in the source file, except for the
byte count.
\end{description}

\section{Examples of using quasi quotes in Prolog}
\label{sec:examples}

In this section we provide three examples to illustrate quasi
quotations. The first example (HTML) is based on the existing SWI-Prolog
libraries for parsing and serializing HTML. The second example
demonstrates how JavaScript can be handled safely (but with limited
syntax checking support) by only tokenising the quotation. The third
example concerns
SPARQL,\footnote{\url{http://www.w3.org/TR/sparql11-overview/}}
demonstrates the value of the quasi quotation approach in a scenario
where the parsed quasi quotation is used as a Prolog query.

\subsection{Quasi quoting HTML}
\label{sec:htmlqq}

The first, complete, example illustrates safe embedding of long HTML
texts into a Prolog web page. The code of the quasi quoter is given in
\figref{htmlqq}. First, we give a fully working webserver based on this
quasi quoter in \figref{date}. We notice that the quoted material can
contain multiple lines, does not require any line endings and may
contain quotes. The content of the quoted material is valid HTML and
because an editor can easily detect the \verb"{|html" and \verb"|}"
indicators, it is not hard for development tools to provide support for
the embedded HTML, such as highlighting, indentation or completion.

\begin{figure}
\begin{code}
:- use_module(library(http/thread_httpd)).
:- use_module(library(http/http_dispatch)).
:- use_module(library(http/html_write)).

server(Port) :- http_server(http_dispatch, [port(Port)]).

:- http_handler(/, clock, []).

clock(_Request) :-
        get_time(Now),
        format_time(atom(Date), '%+', Now),
        reply_html_page(
            title('My digital clock'),
            {|html(Date)||
             <h1>My digital clock</h1>

             <p>It is now <span class="time">Date</span>
             |}).
\end{code}

\noindent
    \caption{Example web server with embedded HTML}
    \label{fig:date}
\end{figure}

The HTML quoter is defined to limit Prolog variables that are replaced
to those that appear as \emph{arguments} to the \const{html} syntax
indicator (\arg{Date} in \figref{date}). The convention to pass
variables that are subject to replacement explicitly has the following
advantages: (1) it avoids a \emph{singleton variable} warning on
\arg{Date}\footnote{This problem can also be resolved by the quasi
quoter by removing singleton variables that appear inside the result
term of the quasi quoter.}, (2) Prolog clauses have a relatively large
number of variables due to the lack of functional notation which can
avoid variables and destructive assignment which allows for reusing a
variable, (3) it makes the substitution more explicit and (4) it allows
the quoter to check that all intended relacements were made.

The HTML quasi quoter as defined in \figref{htmlqq} replaces Prolog
variables indicated by its arguments (\term{html}{Date}) if they appear
as value for an attribute or content of an element. It performs the
following steps:

\begin{enumerate}
    \item Parse the quoted HTML text using \index{load_html/3}\predref{load_html}{3}.  The option
          \term{max_errors}{0} causes the parser to throw a syntax_error
	  exception and abort on the first error.  As this processing
	  happens while reading the source, the HTML syntax error is
	  reported during compilation and includes line, line position
	  and character count information.
    \item The XML DOM structure is recursively traversed and attributes
	  that have a Prolog variable as value or content that matches
	  a Prolog variable from \arg{Syntax} is replaced by this
	  variable.
\end{enumerate}

When the page is generated (\figref{date}), binding of the variable
\arg{Date} completes the XML DOM structure. This structure is serialised
by \index{reply_html_page/3}\predref{reply_html_page}{3}. The serialisation ensures type safety and the
generation of correct HTML syntax.\footnote{The HTML infrastructure
has a global option to select between HTML and XHTML serialisation,
which implies that HTML in the Prolog source may be serialised as
XHTML to the client.}

\begin{figure}
\begin{code}
:- module(html_quasi_quotations, [ html/4 ]).
:- use_module(library(sgml)).
:- use_module(library(apply)).
:- use_module(library(lists)).
:- use_module(library(quasi_quotations)).

:- quasi_quotation_syntax(html).

html(Content, Vars, Dict, DOM) :-
        include(qq_var(Vars), Dict, QQDict),
        with_quasi_quotation_input(
            Content, In,
            load_html(In, DOM0,
                      [ max_errors(0)
                      ])),
        xml_content(QQDict, DOM0, DOM).

qq_var(Vars, _=Var) :- member(V, Vars), V == Var, !.

xml_content(Dict, [Name], [Var]) :-
        atom(Name),
        memberchk(Name=Var, Dict), !.
xml_content(Dict, Content0, Content) :-
        maplist(xml_content_element(Dict), Content0, Content).

xml_content_element(Dict,
                    element(Tag, Attrs0, Content0),
                    element(Tag, Attrs, Content)) :- !,
        maplist(xml_attribute(Dict), Attrs0, Attrs),
        xml_content(Dict, Content0, Content).
xml_content_element(_, Element, Element).

xml_attribute(Dict, Attr=Name, Attr=Var) :-
        memberchk(Name=Var, Dict), !.
xml_attribute(_, Attr, Attr).
\end{code}

\noindent
    \caption{Source for the HTML quasi quoter}
    \label{fig:htmlqq}
\end{figure}

\subsection{Embedding JavaScript}
\label{sec:jsqq}

The HTML quasi quoter of the previous section was easily implemented
because the SWI-Prolog infrastructure already contains a parser
and serializer for HTML. We do not have these for JavaScript.  We are
likely to develop this in the future, but here we want to illustrate
that it is possible to achieve safe template replacement by only using
a \jargon{tokeniser}.

Because JavaScript is generated as part of the HTML page generation,
the JavaScript quasi quoter produces output for the HTML backend.  It
translates the JavaScript into a list of two types of elements: (1)
plain atoms (that will be emitted in the context of a \const{script}
element) and calls to a grammar \index{js_expression//1}\dcgref{js_expression}{1}, which defines a
translation of native Prolog data into JavaScript literals according
to \tabref{jsconvert}.

\begin{table}
\begin{center}
\begin{tabular}{ll}
\hline
\bf Prolog & \bf JavaScript \\
\hline
number	   & number \\
atom	   & string (escaped using JavaScript syntax) \\
@true	   & boolean true \\
@false	   & boolean false \\
@null	   & null constant \\
List	   & array \\
object(NameValueList) & object literal \\
\{ Name:Value, ...\} & object literal \\
\hline
\end{tabular}
\end{center}
    \caption{Prolog to JavaScript conversion}
    \label{tab:jsconvert}
\end{table}

The quasi quoter tokenises the quoted material using an ECMAScript
compliant tokeniser, implemented using a Prolog grammar. The quoter
(\figref{jsquoter}) replaces \emph{identifier} tokens that match with a
Prolog variable with a call to \index{js_expression//1}\dcgref{js_expression}{1}, and translates the
remainder into plain atoms. \Figref{jsqq} shows a shortened predicate
from the SWI-Prolog website that applies the JavaScript quasi quoter to
initialise a jQuery widget called \textit{tagit}. The actual predicate
contains a larger configuration object, just `more of the same'. In this
example, the first block is used to compute server URLs and properties
for \arg{Obj} that we need in the remainder. Next, the \index{html//1}\dcgref{html}{1} call
creates the DOM needed for the \textit{tagit} widget. Here, we could
also have used the HTML quasi quoter. The choice is rather arbitrary in
this case because the fragment is short and defines only structure and
data that is passed in from a variable. Next, we see \index{html_requires//1}\dcgref{html_requires}{1},
which ensures that the page head is extended to load the required
JavaScript and CSS resources and finally, there is the script fragment
with embedded Prolog variables. Because Prolog variables are valid
JavaScript identifiers, the fragment contains valid JavaScript syntax.

\begin{figure}
\begin{code}
prolog:doc_object_page_footer(Obj, _Options) -->
  { http_link_to_id(complete_tag, [], Complete),
    http_link_to_id(show_tag, [], OnClick),
    http_link_to_id(remove_tag, [], Remove),
    object_id(Obj, ObjectID),
    object_tags(Obj, Tags),
    atomic_list_concat(Tags, ',', Data)
  },
  html(div(class('user-annotations'),
           input([id(tags), value(Data)]))),
  html_requires(tagit),
  js_script({|javascript(Complete, OnClick, ObjectID, Remove)||
             $(document).ready(function() {
                $("#tags").tagit({
                    autocomplete: { delay: 0.3,
                                    minLength: 1,
                                    source: Complete
                                  },
                    onTagClicked: function(event, ui) {
                      window.location.href = OnClick+"tag="+
                        encodeURIComponent(ui.tagLabel);
                    },
                    beforeTagRemoved: function(event, ui) {
                      $.ajax({ dataType: "json",
                               url: Remove,
                               data: { tag: ui.tagLabel,
                                       obj: ObjectID
                                     }
                             });
                    }
                  });
              });
            |}).
\end{code}

\noindent
    \caption{Shortened code fragment from the SWI-Prolog website that
	     illustrates the embedding of JavaScript for initializing
	     a widget.}
    \label{fig:jsqq}
\end{figure}

\begin{figure}
\begin{code}
javascript(Content, Vars, Dict, \Parts) :-
        include(qq_var(Vars), Dict, QQDict),
        phrase_from_quasi_quotation(
            js(QQDict, Parts), Content).

qq_var(Vars, _=Var) :- member(V, Vars), V == Var, !.

js(Dict, [Pre, Subst|More]) -->
        here(Here0), js_tokens(_), here(Here1),
        js_token(identifier(Name)),
        { memberchk(Name=Var, Dict), !,
          Subst = \js_expression(Var),
          diff_to_atom(Here0, Here1, Pre)
        },
        js(Dict, More).
js(_, [Last]) -->
        string(Codes), \+ [_], !,
        { atom_codes(Last, Codes) }.

js_tokens([]) --> [].
js_tokens([H|T]) --> js_token(H), js_tokens(T).

%!  diff_to_atom(+Start, +End, -Atom)
%
%   True when  Atom  is an atom  that represents the
%   characters between Start and End, where End must
%   be in the tail of the list Start.

diff_to_atom(Start, End, Atom) :-
        diff_list(Start, End, List),
        atom_codes(Atom, List).

diff_list(Start, End, List) :- Start == End, !, List = [].
diff_list([H|Start], End, [H|List]) :-
        diff_list(Start, End, List).

here(Here, Here, Here).
\end{code}

\noindent
    \caption{Partial source for the \const{javascript} syntax quoter.}
    \label{fig:jsquoter}
\end{figure}

\subsection{Embedding SPARQL}
\label{sec:sparqlqq}

SPARQL\footnote{\url{http://www.w3.org/TR/sparql11-overview/}} is the
query language for the semantic web RDF language. In most of this
section, one may replace SPARQL with SQL. ClioPatria is SWI-Prolog's
semantic web framework, which contains a SPARQL `endpoint'. The SPARQL
engine compiles a SPARQL query into a Prolog query, optimizes and
executes this query and serializes the results according to the SPARQL
result specification.

When writing middleware (reasoning) in the ClioPatria, one typically
uses direct queries to the embedded RDF store for the reasoning. In some
cases, one would like to use SPARQL for specifying the query. Consider
cases where the query is already available in SPARQL, the query must
also be used with external servers or the author is much more fluent in
SPARQL than in Prolog. The quasi quotation syntax can be used to write
down the code below. The SPARQL query is parsed at compile time and the
SPARQL projection variables can naturally integrate with the Prolog
variable of the enclosing clause.  Note that in this example, we think
it is better not to pass the substituted variables as arguments to the
\const{sparql} syntax term because this list is made explicit in the
variable projection clause that starts the query.

\begin{code}
   ...,
   {|sparql||
    SELECT ?Name, ?Place WHERE {
            ...
    }
    |}>
\end{code}

\noindent
A similar approach is feasible with SQL, offering a more convenient way
to interact with the database than using \jargon{prepared statements}
and both a safer and more convenient way than direct text-based SQL
queries.

\section{Implementation}
\label{sec:implementation}

The implementation of quasi quotation handling in an existing Prolog
system is straightforward and should not require more than a couple of
days. The tokeniser is extended to recognise \verb"{|". On encountering
this token, the parser builds a list of terms, each of which represents
a quasi quotation. If the list of quasi quotations is not empty when
\index{read/1}\predref{read}{1} reads the fullstop token, it materializes the variable dictionary
and calls a routine that calls the quasi quotation parsers. Note that
quasi quotations are parsed \emph{after} \index{read/1}\predref{read}{1} completes reading the
term. This allows for quasi quotations to refer to variables that appear
\emph{after} the quasi quotation (see \secref{sparqlqq}). This does
imply that \index{read/1}\predref{read}{1} must buffer the quoted text. The functionality is made
available to the user through the library \const{quasi_quotations.pl},
which appeared in SWI-Prolog 6.3.17.

The Prolog tokeniser is extended with two tokens. \verb"{|" is a new
token that starts a quasi quotation. The sequence \verb"||...|}" is
processed as a single token.

\section{Future work}

Quasi quotations are young in SWI-Prolog. We believe that the core
functionality described in this article is largely future-proof.

One of the future tasks is to establish libraries that facilitate syntax
handling and safe replacement.  Another is to establish design patterns
for using this technology.  What can be learn from the Haskell, Python
and Ruby communities here?  Quasi quotations typically `evaluate' to
a Prolog term that shares variables with the clause into which it is
embedded.  We do not have a functional syntax that we can exploit to
force `evaluation' of this term with instantiated variables.  On the
other hand, we can also use the non-ground abstract syntax tree and
use it for e.g., \textit{matching} tasks.

\section{Conclusions}

In this article we have motivated why it is necessary to be able to
embed longer snippets of code written in another language in Prolog
source code. This requires for a syntax that permits embedding,
practically without the need for Prolog escaping in the embedded text.
Next, this text must be related to Prolog data. The integration of
Prolog data into the snippet must be done according to the syntax and
data model of the external language. This implies we need a programmable
component that is related to the quoted material.

Combining these requirements into one syntactic extension that is
executed by \index{read/1}\predref{read}{1} simplifies support by tooling such as editors,
provides natural access to Prolog variables and allows for seemless
integration of error messages.

We have demonstrated quasi quotations using two implemented quoters.
These quoters allow for easy copying and pasting material in their
native syntax to and from Prolog source.  Using these quoters is
likely to reduce the learning curve for embedding snippets into
Prolog, while the quasi quoter can guarantee that the integrated
material is (syntactically) correct and that Prolog material is
\emph{safely} integrated.

%================================================================
\subsection*{Acknowledgements}

Jan Pobrislo provided the insights that were used to write
\secref{thebigthree}. Michiel Hildebrand and Jacco van Ossenbruggen (VU
University Amsterdam) have tried many of the pure Prolog based
alternatives which were needed to formulate \secref{needqq} and
\secref{syntaxext}. They also help shaping the current JavaScript quasi
quoter and commented on drafts of this text. Ulrich Neumerkel has
propose the combination of \chr{\Sbar} with brackets and pointed at the
tokenisation issues described in \secref{implementation}.

This publication was supported by the Dutch national program COMMIT/

\bibliographystyle{plain}
\bibliography{quasiquoting}

\egroup

\end{document}